\documentstyle[12pt]{article}

\begin{document}
\begin{titlepage}
\centerline {\bf Saturation properties and incompressibility of nuclear
matter:}
\centerline {\bf  A consistent determination from nuclear masses}
\medskip
\centerline {R. Nayak}
\centerline {\it Physics Department, Ravenshaw College, Cuttack, India}
\centerline {\it and Laboratoire de Physique Nucleaire,Universite de Montreal,
             Montreal,H3C 3J7 Canada}
\smallskip
\centerline {V.S. Uma Maheswari and L. Satpathy}
\centerline {\it Institute of Physics, Bhubaneswar-751 005, India}
\smallskip
\begin{abstract}
{ Starting with a two-body effective nucleon-nucleon interaction, it
is shown that the infinite nuclear matter model of atomic nuclei is
more appropriate than
the conventional Bethe-Weizsacker like mass formulae to extract
saturation properties of nuclear matter from nuclear masses. In
particular, the saturation density thus obtained agrees with that of
electron scattering data and the Hartree-Fock calculations. For the
first time using nuclear mass formula, the radius constant
$r_0$=1.138 fm  and binding energy per nucleon $a_v$ = -16.11 MeV,
corresponding to the infinite nuclear matter, are consistently
obtained from the same source.
An important offshoot of this study is the determination of nuclear
matter incompressibility $K_{\infty}$ to be 288$\pm$ 28 MeV
using the same source of  nuclear masses as input.}

\end{abstract}
\end{titlepage}
\newpage
\centerline {\bf I. INTRODUCTION}
\vskip 1.cm
The binding energy, saturation density and compression modulus of infinite
nuclear matter are fundamental constants of nature.
Traditionally, the first two quantities termed as saturation properties
are determined from two different sources, namely the volume coefficient
$a_v$ of the Bethe-Weizsacker (BW) like mass formulae and the
electron scattering data on heavy nuclei respectively.
Although the Coulomb coefficient $a_C(=0.6e^2/r_0)$ in BW-like mass
formulae specifies the density $\rho=3/(4\pi r_0^3)$, it is not
accepted as the density of nuclear matter. This is because, the
corresponding radius constant $r_0 \simeq 1.22 fm$
obtained\cite{myers,sh75} in a totally free fit is much higher than the
value $1.12-1.13$ fm obtained from the electron scattering
data\cite{deshalit} on heavy nuclei and Hartree-Fock
calculations\cite{gog} .
As yet no mass formula fit to nuclear masses has yielded a value of
$r_0$ in this range. This is the so-called ``$r_0 - $\ paradox",
which has been a subject of investigation\cite{anamoly} over the
years by many. Since, the
 two properties are highly inter-related, the above constrained practice
of their determination from two different sources, has been a serious
discomfeature in our understanding of nuclear dynamics.
Coupled to this, the
incompressibility of nuclear matter has posed a much serious problem
with regard to its determination, both theoretically and experimentally.

In this work, we report our attempt to determine all the three
properties of nuclear matter using a single model,
and one kind of experimental data, namely the nuclear masses, which
are abundant in nature and are the best known properties of nuclei. We use
the
infinite nuclear matter (INM) model\cite{inm1} based on the generalised
Hugenholtz-Van Hove (HVH) theorem\cite{inm2} of many-body theory, whose
success
has been well tested through its unique ability to predict masses of
nuclei far from stability\cite{inm1}, masses of $Na$ isotopes
 and other light nuclei,
 and finally through the 1986-87 mass predictions\cite{inm4} of the
entire periodic table. In the formulation of INM model, it was
claimed\cite{inm1} that this model is more suitable than the
traditional (BW) ones
to extract the properties of nuclear matter, as it is exclusively
built in terms of infinite nuclear matter at ground-state. In the
present work, we have improved the model and show conclusively,
starting from two-body effective interaction within the energy
density formalism,  that the
saturation properties derived through this improved model are closer
to the true properties of the nuclear matter than those derived using
BW model based  mass formulae. Then this model is  fitted to  the
experimental masses, which yields a value of $r_0$ to be
 1.138 fm, in close agreement with that obtained from the
electron scattering data, and with the empirical value found
through many-body mean-field approaches\cite{jmp2}.
Further, using these saturation properties determined from the same
set of data  on nuclear
masses, we arrive at a  value of about 288 MeV  for the
incompressibility $K_{\infty}$, which is  of equal
fundamental importance in the realm of nuclear physics and astrophysics.

In Sec. II, the improvements we have made in the INM model are
presented. In Sec. III, we show at a microscopic level that the
improved INM model is more appropriate than the BW-like model for the
extraction of saturation properties of infinite nuclear matter from
the nuclear masses. Determination of such properties are presented in
Sec. IV. In Sec.V, the value of nuclear compression modulus is determined
from the nuclear masses. Finally, we conclude in Sec.VI.

\centerline {\bf II. THE IMPROVED INM MODEL}

We recall here the essential features
of the INM model\cite{inm1} which we have now improved.
In this model, the ground-state energy $E^F(A,Z)$ of a
nucleus(A,N,Z) with asymmetry $\beta$ is considered equivalent to the
energy $E^S$ of a perfect sphere made up of infinite nuclear matter
at ground-state density with same asymmetry $\beta$
plus the residual energy $\eta$, called the local energy, which
contains all the characteristic contributions like shell, deformation
etc. So,
\begin{equation}
E^F(A,Z) = E^S_{INM} (A,Z) + \eta (A,Z)
\end{equation}
with $E^S_{INM}(A,Z) = E(A,Z)+f(A,Z)$, where
\begin{equation}
f(A,Z) =
a_s^IA^{2/3}+a_C^I(Z^2-5({3/ (16\pi)})^{2/3}Z^{4/3})A^{-1/3}
+a_{ss}^IA^{2/3}\beta^2+
a_{cv}^IA^{1/3}-\delta(A,Z)
\end{equation}
denotes the finite-size effects and $E(A,Z)$ is the energy of the
infinite part. The superscript $I$ refers to the
INM character of the coefficients.
Here $a_s^I$, $a_C^I$, $a_{ss}^I$ and $a_{cv}^I$ are the
surface, Coulomb, surface-symmetry and curvature coefficients.
$\delta(A,Z)$ is the usual pairing term, given as
\begin{eqnarray}
\delta (A,Z) &=& + \Delta A^{-1/2} \quad {\rm for\ even-even\ nuclei}
\nonumber \\
&=& 0 \quad {\rm for\ odd-A\ nuclei} \nonumber \\
&=& - \Delta A^{-1/2} \quad {\rm for\ odd-odd\ nuclei} \nonumber
\end{eqnarray}
Eq.(1) now becomes,
\begin{equation}
E^F(A,Z) = E(A,Z) + f(A,Z) + \eta (A,Z)
\end{equation}
Thus, the energy of a finite nucleus is given as a sum of three
distinct parts; an infinite part $E(A,Z)$, a
finite-size component $f(A,Z)$ and a local energy part
$\eta(A,Z)$.
The term $E(A,Z)$ being the property of infinite nuclear
matter at ground-state, will satisfy the generalised HVH
theorem\cite{inm2}.
\begin{equation}
{E/ A} =  [
(1+\beta)\epsilon_{n}+(1-\beta)\epsilon_{p} ]/2
\end{equation}
where $\epsilon_n=(\partial E/\partial N)_Z$ and
$\epsilon_p = (\partial E/\partial Z)_N$ are the
neutron and proton Fermi energies respectively. Using Eq.(3), the INM
Fermi energies $\epsilon_n$ and $\epsilon_p$ can be expressed in
terms of their counterparts for finite nuclei as
\begin{equation}
\epsilon_n= \epsilon_n^F-{(\partial f/\partial N)}\mid_Z
-{(\partial \eta /\partial N)}\mid_Z ;
\epsilon_p= \epsilon_p^F-{(\partial f/\partial Z)}\mid_N
-{(\partial \eta /\partial Z)}\mid_N
\end{equation}
where $\epsilon_n^F=(\partial E^F/\partial N)_Z$
and $\epsilon_p^F=(\partial E^F/\partial Z)_N$.
Using (3) and (5), Eq.(4) is rewritten as
\begin{equation}
{E^F/ A} =  [
(1+\beta)\epsilon_{n}^F+(1-\beta)\epsilon_{p}^F ]/2 + S(A,Z)
\end{equation}
where,
$S(A,Z)= {f/ A}-{(N/ A)}{(\partial f/ \partial N)}_Z
-{(Z/ A)}{(\partial f/ \partial Z)}_N $
is a function of all the finite-size terms $a_s^I$, $a_C^I$,
$a_{ss}^I$ and $a_{cv}^I$, which are global in nature.
As discussed earlier\cite{inm1}, the $\eta$ terms in
Eq.(6) drops out,
which plays a crucial role in the success of INM model, and whose
validity has been amply demonstrated\cite{inm1,inm4}.
It must be noted that Eq.(6) does not contain the infinite part $E$
as well as the $\eta$ terms.
Thus, through Eq.(6), the decoupling of the finite
component $f$ from the infinite one $E$ has been acheived. The
coefficients $a_s^I$, $a_C^I$, $a_{ss}^I$ and $a_{cv}^I$ can be
determined by fitting $S(A,Z)$ function with the combination of data
${E^F/ A} - [(1+\beta)\epsilon_{n}^F+(1-\beta)\epsilon_{p}^F]/2$
obtained from the nuclear masses. We would like to mention here that
in the earlier work\cite{inm1}, due to the use of the expressions for
Fermi energies, $\epsilon_n^F=E^F(N,Z)-E^F(N-1,Z)$ and
$\epsilon_p^F=E^F(N,Z)-E^F(N,Z-1)$, a small contribution
$a_{a}(\beta^2-1)/(A-1)$ survives (of the order of
$a_{a}/A$) in Eq.(6), whereas in the present work by using the
better formulae
\begin{eqnarray}
\epsilon^F_n &=& {\partial E^F\over \partial N}\mid_Z =
{1\over 2} \left[ E^F(A+1,Z)-E^F(A-1,Z)\right] \nonumber \\
\epsilon^F_p &=& {\partial E^F\over \partial Z}\mid_N =
{1\over 2} \left[ E^F(A+1,Z+1)-E^F(A-1,Z-1)\right]
\end{eqnarray}
the following important improvements are acheived.

\begin{description}
\item (i). The decoupling of the infinite part (asymmetry term) from the
finite part in Eq.(6) occurs upto an order of $a_{a}/A^2$, which can
be considered perfect at the numerical level.
\item  (ii). The pairing term $\delta$ contained in $f$ effectively
drops out in Eq.(6), thereby rendering the determination of other
coefficients with greater accuracy due to less correlation.
\item  (iii). The exchange Coulomb term of the
standard form $O(Z^{4/3}A^{-1/3})$ (2) exactly cancels in Eq.(6).
This cancellation gives rise to a more reliable determination of INM
saturation density through $a_C^I$.
\item (iv) The other factors which
might affect the determination of density such as proton-form-factor
[$O(Z^2/A)$] and Nolen-Schiffer anomaly [$O(\beta A)$,
referred to as charge-asymmetry energy] also
cancel exactly.
\end{description}
Thus all the finite-size
coefficients contained in $S(A,Z)$, which are global in nature, are
determined from nuclear masses by a fit to Eq.(6).
Now, of the three distinct parts of the energy $E^F$ of a finite
nucleus(3), the infinite part $E$ and the local energy part $\eta$
remain to be determined.
The infinite part $E$ must satisfy the generalised HVH theorem (4),
whose solution is of the form,
\begin{equation}
E=-a_v^I A + a_{a}^I \beta^2 A
\end{equation}
where $a_v^I$ and $a_{a}^I$ are the global parameters which can
be identified as volume and symmetry coefficients corresponding to INM.
Using (5) and (8) in the right and left hand sides of Eq.(4) respectively,
one obtains
\begin{equation}
-a_v^I+a_{a}^I \beta^2 =
{1\over 2} \left [ (1+\beta)\epsilon_{n}^F+(1-\beta)\epsilon_{p}^F \right ]
-\left [ {N\over A}{\partial f\over \partial N}\mid_Z+
{Z\over A}{\partial f\over \partial Z}\mid_N \right ]
\end{equation}
where the contribution from the local energy part( of the
order of $\eta/A$) is neglected, which in the limit of large A goes
to zero.
Since $f$ is known from Eq.(6), the above equation can be used to
determine the two parameters $a_v^I$ and $a_{a}^I$ with the
combination of data
$\left [ (1+\beta)\epsilon_{n}^F+(1-\beta)\epsilon_{p}^F \right ]/2$
obtained from nuclear masses. Thus all the global parameters are determined
essentially in two fits: Eq.(6) determines the finite-size
coefficients like $a_s^I$, $a_C^I$ {\it etc} and Eq.(9) determines
the INM coefficients $a_v^I$ and $a_{a}^I$.
Since the present study is intended for the determination of the
properties of nuclear matter, we do not discuss the determination of
$\eta$ and consequently the masses, the details of which can be seen
in Refs.\cite{inm1,inm4}

\centerline  {\bf III. IMPROVED INM MODEL VERSUS BW MODEL}

In this section, we would like to make a comparative study of the
improved INM model and BW model, in regard to their suitability for
the determination of saturation properties of infinite nuclear matter
from nuclear masses.

As noted in the introduction, it has not been possible to determine
both the energy and density of infinite nuclear matter in the BW
model based mass formulae. Further, it has been hoped only that the
volume coefficient determined in the BW model corresponds to nuclear
matter at ground state. On the other hand, in the INM model, this
fact has been ensured by the explicit use of HVH theorem, which is
valid only at the ground-state of infinite nuclear matter. Since in
this model, the binding energy of a nucleus is written in terms of
the properties of INM, it is expected that the INM model is well
equipped to extract infinite nuclear matter properties from nuclear
masses. We demonstrate this  by predicting
the {\it a}\ {\it priori} known INM properties for a
given effective interaction.

In this regard, we make use of the extended
Thomas-Fermi(ETF) calculation\cite{etf} of nuclear binding energies with
Skyrme-like forces, which over the years has been firmly established.
In such calculations, one obtains the
smooth part of the energy corresponding to the liquid-drop nature of
nuclei. This smooth part, hereafter referred to as macroscopic part
describes the energy $E_{INM}^S$ of the INM sphere  as defined in
Eq.(1). For the
purpose of making a comparative study of the INM model and the
BW-like model, the appropriate BW mass formula is
\begin{eqnarray}
E_{BW} &=& -a_v^L A + a_s^L A^{2/3} +
a_C^L(Z^2-5({3\over 16\pi})^{2/3}Z^{4/3})A^{-1/3}
+ a_{a}^L \beta^2 A  \nonumber \\
& & +a_{ss}^L A^{2/3}{\beta }^2 + a_{cv}^L A^{1/3} - \delta (A,Z)
\end{eqnarray}
In the case of the ETF calculations, nuclear
curvature coefficient comes out to be about 10 MeV as against the BW-like
mass formula fit to real nuclei, which gives a value close to zero.
For this reason, we have included higher-order terms like curvature
and surface-symmetry terms in both the INM and BW models.

The macroscopic or the ETF nuclear ground state energies used here for
the comparative study of the INM and BW models are taken from
the calculations of Aboussir {\it et al}\cite{jmp}.
In their calculation, they used a generalised Skyrme force SkSC4 of
the form
\begin{eqnarray}
{v_{ij}} &=& t_0 [1+x_0P_{\sigma }\delta ({\vec r_{ij}})] \nonumber \\
\quad   &+& t_1 ( 1+x_1P_{\sigma}) [p_{ij}^2 \delta ({\vec r_{ij}})^2
+h.a.]/2\hbar^2 \nonumber \\
\quad   &+& t_2 ( 1+x_2P_{\sigma}){\vec p_{ij}} \cdot \delta ({\vec
r_{ij}}) {\vec p_{ij}}/\hbar^2 \nonumber \\
\quad   &+& (t_3/6) ( 1+x_3P_{\sigma})[\rho_{qi}({\vec
r_i})+\rho_{qj}({\vec r_j}) ]^{\gamma} \delta ({\vec r_{ij}})
\nonumber \\
\quad   &+& (i/\hbar^2 )\ W_0 ({\vec \sigma_i} + {\vec \sigma_j } )
\cdot {\vec p_{ij}} \times \delta ({\vec
r_{ij}}) {\vec p_{ij}}. \nonumber
\end{eqnarray}
Then, the macroscopic part of the total energy for a given nucleus
is calculated using the energy density formalism, i.e. $E = \int
{\cal E} ({\vec r}) d^3r $, where ${\cal E} = \tau ({\vec r}) +
v({\vec r}) $. The potential energy density $v$ is derived using the
two-body force given above. For the kinetic part $\tau $, they use the full
fourth order ETF kinetic functional\cite{etf}. It must be noted
that realistic nuclear ground state energies contain shell effects.
To incorporate this characteristic feature in a self-consistent way,
a Hartree-Fock calculation is performed for the same generalised
Skyrme force. Now, using the single particle states obtained within
the HF approximation, the shell corrections can be calculated by directly
making use of the Strutinsky procedure. Then, the total energy is
given as the sum of the macroscopic part and these shell corrections.

We have made
an exhaustive study using those macroscopic energies provided by them
for 1492 nuclei.
We fitted both the INM formulae given by (6) and (9)
and the BW one by (10) to the above macroscopic part of the nuclear
masses to determine the corresponding global paramters. The results
so obtained, together with their respective errors, are given in Table I.
The values obtained directly by Aboussir {\it et al} with the SkSC4,
force performing nuclear matter
and semi-infinite nuclear matter calculations for
the various coefficients (hereafter referred to as exact values)
are also presented in Table I.
It is gratifying to find that the values obtained in the INM fit for
the principal coefficients like $a_v^I$, $a_s^I$ and $a_C^I$ agree
better with the exact  values, compared to that of the BW fit. The
symmetry coefficient $a_{\beta}^I$ agrees reasonably well with the
exact  value, although somewhat inferior to the BW value. Even though the
agreement of the higher-order terms like surface-symmetry and
curvature in the BW-fit agree better, it must be noted that, because
of correlations amongst the coefficients, they significantly affect the
principal term like surface, and to a lesser extent the other ones also.
In case of the INM fit, since the infinite and finite parts are
determined in two separate fits, the principle coefficients are not
influenced by the higher-order terms. In any case, these two
coefficients contribute insignificantly in real nuclei, and are
normally ignored.
Thus, the saturation properties of nuclear matter, which are {\it a}
{\it priori} known for a given effective interaction like SkSC4 is
relatively well reproduced by the INM model than the BW-like model.
This gives us more confidence in the INM model in extracting
real saturation properties from experimental nuclear masses which is
done in the next section.

The success of the INM model over that of the BW-like model is
essentially due to the following.
As also discussed in Ref.\cite{inm1}, the BW-like mass
formulae use only the average property of nuclear matter, namely the
average energy per nucleon.
However, as demonstrated by Hugenholtz-Van Hove\cite{hvh}, an
interacting Fermi
system has an additional property namely the single-particle
property. Such a system has one true single-particle state {\it i.e.}
Fermi state, which has infinite life time, while other low lying ones
are metastable.
In other words, the lifetime of the single partcle state approaches
infinity in the limit $k \longrightarrow k_F$.
This important property is additionally taken into
account in the INM model, which is not present in BW-like mass formulae.

\centerline   {\bf IV. DETERMINATION OF NUCLEAR MATTER SATURATION PROPERTIES}

Before coming to the actual determination of the various parameters
of the INM model, and thereby the saturation properties of nuclear
matter, it is essential to assess the relative
importance of the possible higher-order terms, which is somewhat
different in this model.

The two saturation properties, namely the density $\rho_{\infty}$
given by $a_C^I$ and the volume energy $a_v^I$ are determined in two
different fits; Eqs.(6) $\&$ (9) respectively. The first fit
determines the crucial quantity $a_C^I$; and hence, it is imperative
that we analyse the role of other finite-size effects
which may influence the determination of the saturation properties.
The finite-size terms which are directly related to the Coulomb
effect are: exchange Coulomb, proton-form-factor correction and
charge-asymmetry energy.  It may be recalled here that in the INM
model, the binding energies and Fermi energies are used in the
particular combination, $E^F/ A =  [
(1+\beta)\epsilon_{n}^F+(1-\beta)\epsilon_{p}^F ]/2  $, in Eq.(6), as
dictated by the HVH theorem. As a result, the above stated three
effects exactly cancel in Eq.(6) rendering a clean determination of
$a_C^I$, and hence the density $\rho_{\infty}$. This is indeed a very
fortunate situation.

The other two higher-order terms which may indirectly affect the
value of $a_C^I$ are the curvature $a_{cv}^I$ and
surface-symmetry$a_{ss}^I$  coefficients.
In real nuclei, the curvature coefficient comes out to be nearly
zero, and is normally not included. So, we have dropped it.
In regard to the surface-symmetry coefficient $a_{ss}^I$, it has been
recoganised that its value is somewhat difficult to determine from
nuclear masses. Even at the theoretical level, the value of
$a_{ss}^I$ determined\cite{etf} from various effective interactions differ
widely. Further, in the modern BW-like mass formulae\cite{Mn87,My87}, this
coefficient is fixed from considerations other than the ground-state
nuclear masses, such as fission barrier heights. Since, in the
present study, we address ourselves to the determination of the
properties of INM at ground-state, it is essential that we only use
the ground-state masses, and not any other property which may drift
the system from the ground-state and jeopardize the determination of
$a_v^I$ and $a_C^I$. Therefore, in the present context, it is proper
that the important coefficients are treated as free parameters to be
fixed by nuclear masses through Eq.(6).

But, fortunately, the surface-symmetry coefficient $a_{ss}^I$ cancels
to a major extent $(\sim
66 \% )$ due to the particular combination of data used in Eq.(6).
Although this
does not fully cancels like the exchange Coulomb, protron-form-factor
correction terms etc, $a_{ss}^I$ being a second-order term, such
cancellation renders it relatively insignificant as compared to
$a_s^I$. At numerical level, it may be considered to
be virtually cancelled. Nevertheless, since our main goal is to determine the
saturation properties of nuclear matter, which are of fundamental
importance, we are anxious to check if
any semblance of survival of $a_{ss}^I$ term can affect the results.

Hence, we carried out calculations retaining this term as a free
paramter in our fit to Eq.(6).
It is found that while the values of the other coefficients remain
almost unaltered, the value of $a_{ss}^I$ widely varies from -30 MeV
to -11 MeV with accompanied error of about 50-100$\%$ as number of
data varies from 1085 to 1371. This fact is also true when one uses
presuppossed values for $a_{ss}^I$, while other coefficients are
being fitted to the 1371 masses. As the value of $a_{ss}^I$ is varied
from -10 MeV to -30 MeV, it was found that $\chi_{rms}$ shows a
minimum at $a_{ss}^I \sim -12 $MeV. However, this optimum
value of $a_{ss}^I$ fluctuates with the variation of the number of
data, resulting in no definite value.
These two features are reminiscent of its insignificant presence in
Eq.(6). Hence we have omitted this term. The same is true also for the
curvature term.
Therefore, the optimum representation for the finite-size function
$f(A)$ defined in Eq.(1) is $f(A,Z) =
a_s^IA^{2/3}+a_C^I(Z^2-5({3/ (16\pi)})^{2/3}Z^{4/3})A^{-1/3}$.

Now coming to the actual determination of  the saturation properties of
INM,  we use all the nuclear masses
with experimental error $\le$ 60 keV from the recent mass table of Wapstra and
Audi\cite{audi}. There are 1371 cases, which have been used in our study.
As mentioned earlier the universal parameters in this model are determined
in a two-step process. In the first step, we determine the finite-size
coefficients $a_C^I$
and $a_s^I$ by making a least-square fit to Eq.(6) using all the 1371
masses. Then these parameters so determined
 are further used in the second step to obtain
 the coefficients corresponding to the infinite
part $a_v^I$ and $a_{\beta}^I$,  by a fit to Eq.(9)
using the same set of data. The $\chi_{rms}$ obtained for these two fits are
371 keV and 372 keV respectively, which are substantially lower than
the corresponding ones 460 keV and 506 keV obtained in the earlier
study\cite{inm1,inm4}.
 The lowering of $\chi_{rms}$ is almost entirely due to the
improvements made in the
model, and  not as the result of the use of recent masses.
To check the goodness and the stability of the parameters obtained in
our fits, we have carried out
five sets of calculations by varying the number of data randomly considered
through out the mass table, choosing them on the basis of experimental
error ranging from 20 keV to 60 keV, and these  are
presented in Table II. One can clearly see that almost all the the four
parameters are
quite stable inspite of widely varying data. Especially remarkable is the
stability of the two crucial nuclear  parameters
namely , the Coulomb coefficient $a_C^I$, and
and the volume coefficient   $a_v^I$.
It is satisfying to note that the degree of stability of these two
important parameters, in which we are specifically interested,
is relatively better than in $a_s^I$ and $a_{\beta}^I$.
The final values obtained for
these two coefficients with maximum number of data (1371 nuclei), and the
corresponding values for $r_0$ and $\rho_{\infty}$ are:
\begin{eqnarray}
a_v^I = 16.108 {\rm MeV} \quad & {\rm and} & \quad a_C^I = 0.7593
{\rm MeV} \nonumber \\
r_0 = 1.138 {\rm fm} \quad & {\rm and} & \quad \rho_{\infty} = 0.1620
{\rm fm^{-3}} \nonumber.
\end{eqnarray}
We quote no errors for our parameters as they are firmly determined,
say  with errors less than  $1\%$.
The saturation properties $a_v^I = 18.335$ MeV and $a_C^I = 0.841$ MeV,
determined earlier\cite{inm1} are inaccurate due to the use of the expressions
$\epsilon_N^F = E^F(N,Z)-E^F(N-1,Z)$ and $\epsilon_P^F = E^F(N,Z)-E^F(N,Z-1)$
for finite nuclei Fermi energies.
It is indeed remarkable that the saturation
density$\rho_{\infty} = 0.162 fm^{-3}$ and the
corresponding $r_0 = 1.138 fm$ found here agree quite well with
that obtained from the fit of  electron scattering data. This value
of $r_0$ is also close to 1.13 fm  obtained in the HF studies,
which has been widely accepted in literature\cite{gog}.
It may be noted that our value of $r_0$ is quite similar to the value
1.140$\pm$.005  obtained from the fit of nuclear charge
radii\cite{jmp2} extracted from the recent electron scattering
data\cite{vri}.
Thus, the two important
ground-state properties,{\it i.e} $a_v$ and $\rho_{\infty}$
which are inter-related, are consistently determined from one kind of
data using a single model.

\centerline {\bf V. INCOMPRESSIBILITY OF NUCLEAR MATTER}

To determine $K_{\infty}$, we note that INM model determines binding
energy per nucleon $a_v$
and saturation density $\rho_{\infty}$ at ground-state. The BW model
based
mass formulae give the value of $a_v^L$ at a different density
$\rho_o$ corresponding to their $a_c^L$, since they do not have any
ingredient to ensure that these parameters pertain to nuclear matter
at ground-state. Hence, using the values of $a_v$ and densities from
INM model as well as from BW formula, one can determine $K_{\infty}$
using the relation
\begin{equation}
a_v^I(\rho_{\infty}) + {(K_{\infty}/ 18)}
({\rho_{o}/\rho_{\infty}-1 )}^2 = a_v^L(\rho_{o})
\end{equation}
as shown in Ref.\cite{lscomp}.

In order to determine the optimum number of parameters in the BW mass
formula given by Eq.(10), we have carried out least-squares fit with
varying number of parameters, the results of which are presented in
Table III. We have used the same 1371 nuclear masses mentioned in our
earlier section. It can be seen that the values of the principal five
coefficients are not affected when the surface-symmetry $a_{ss}^L$
and curvature $a_{cv}^L$ terms are successively included. Hence, the
$a_{ss}^L$ term is well supported and should be retained. The
curvature term, in spite of
its smallness and relatively large error, can be included as it does
not affect the leading terms much. However, the inclusion of the
Gauss curvature $a_{gc}^L$ term, the next higher-order term
in the model, completely destabilises the fit by violently disturbing
the leading order coefficients. The surface coefficient has even
become negative.
This may be due to the very small value\cite{Mn87,My87} of this term
$a_{gc}^L$, which is of the order of 6 MeV.

Quite importantly, the above result is contrary to the common belief
that the inclusion of more and more
higher-order terms in a liquid-drop model like expansion would result
in progressively refined values of the leading order
terms. Therefore, one should be judicious in retaining higher-order
terms in such models. In the present study, we consider
Eq.(10) having six parameters to be the optimum representation, where we
have dropped the curvature term as it comes out to be nearly zero.

With this view, we carried out a least-square fit to Eq.(10) (without
the curvature term) using the same 1371 masses.
As in the case of INM
model calculation, we have varied the number of data to arrive at
stable values of $a_v^L$ and $a_C^L$, with similar accuracies of second
and third decimal places respectively,
since the value of $K_{\infty}$ is sensitive to these values.
 The results are given in Table IV.
Now, $K_{\infty}$ is computed using these values in Eq.(11) together
with the values of $a_v^I$ and $a_C^I$ from Table II for the
corresponding set of data, and are presented in the last column of
Table IV. It is remarkable that inspite of variation of the number of
input data ranging from 1085 to 1371, the value of
 $K_{\infty}$ comes out to be in between 288 and 305.
The average value thus obtained in Table IV is about 294 MeV which is very
close to 288 MeV obtained with the maximum number of data used in the fitting
procedure, which futher substantiates the stability of our result
with respect to the variation of data.

We then attempt to get an estimate of the error in this value of
$K_{\infty}$ arising out of the limitations of the model, which may
be due to the inclusion/non-inclusion of higher-order
terms like curvature, exchange Coulomb and proton-form-factor. The
results of our calculation
of $K_{\infty}$ with inclusion/non-inclusion of these three effects
are presented in Table V.  We have calculated
the error $\chi$ using the expression $\chi^2 = {1\over N} {\Sigma_{i=1}^{N}}
{\left ( K_{\infty}^i - K_{\infty}^{opt} \right )}^2$, where
$K_{\infty}^{opt}$ = 288 MeV, $N=5$ and $i$ stands for the five
values, other than $K_{\infty}^{opt}$, tabulated in Table V.  The
error thus calculated comes out to be 28 MeV.

The recent BW model based  mass formulae usually use presupposed
value of $r_0$ determined from other considerations. The one which treats
$r_0$ as an adjustable parameter and more or less looks similar to Eq.(10)
is by Myers and Swiatecki\cite{myers}, where $r_0$ is determined by
using the data on both the
nuclear masses and fission barriers. Using their values of $a_v$ and
the density, and the present values of INM, we obtain $K_{\infty}$ to be about
 299 MeV.
Hence, we would like to firmly state that, if one allows $r_0$ as a free
parameter in the fit to nuclear masses, one would invariably arrive at
a value of about 288$\pm$ 28 MeV for $K_{\infty}$.

\centerline {VI. CONCLUSIONS}

In conclusion, we have improved the INM model by using better Fermi
energies for neutron and proton, which has resulted in a cleaner
decoupling of the finite-size effects and the INM part of
the ground-state energies of nuclei. Unlike in the BW-like mass
formulae, the Coulomb related higher-order terms such as exchange
Coulomb, proton-form-factor correction and charge-asymmetry energy
cancel exactly, rendering accurate determination of the most
important quantity, namely the saturation density.
More importantly, we have demonstrated at a fundamental level
starting with effective two-body interaction, the
appropriateness of the INM model over that of the BW-like models to
determine the ground-state properties of INM.

The saturation density $\rho_{\infty}$ and binding energy per nucleon
$a_v$ of nuclear matter, the two highly inter-related quantities, are
extracted consistently for the first time from a single source, i.e.
nuclear masses, through a mass formula.
 It is particularly satisfying to find that the radius
constant corresponding to $\rho_{\infty}$ determined here agrees
quite well with that obtained from
electron scattering data and Hartree-Fock calculations.
These have been possible because of
taking into account additionally the single particle property of the
interacting Fermi system through the use of the generalised
HVH theorem in the INM model. Thus, the $r_0 - $ anomaly is resolved
here satisfactorily.

An important offshoot of this study is the determination of the value of
nuclear matter incompressibility starting from nuclear masses, which are the
best measured and
most abundant data in nuclear physics. The value so obtained for  $K_{\infty}$
is  288$\pm$ 28 MeV. We finally commement that inclusion of
suface-symmetry term $a_ss^{I}$ leads to fluctuation of the value of
$K_{\infty}$ to a larger side of the above value.

\centerline {\bf ACKNOWLEDGMENTS}

We are greatly indebted to Professor J.M. Pearson for
providing us the ETF masses. We are also thankful to him for very useful
discussions and critical reading of the manuscript. Helpful
discussions with  Professors J.R. Nix and  J.P. Blaizot are
gratefully aknowledged.

\vfill
\newpage
\centerline {\bf Table Captions}

\noindent {\bf Table I.} Values obtained for the global
parameters(Eqs. 6,9-10) in the infinite nuclear matter(INM) model and
Bethe-Weizsacker(BW) mass formula fit using the macroscopic part of
nuclear energies(see text). Exact values determined directly using INM
and semi-INM calculations are also given. All quantities are in MeV.

\noindent {\bf Table II} Values obtained for the global
parameters( Eq. 6,9) in the present study using the experimental
data\cite{audi} on nuclear masses are given for the various sets of data.
All quantities are in MeV.

\noindent {\bf Table III} Values obtained for the parameters(Eq. 10) of BW
model using the experimental
data\cite{audi} with varying number of higher order terms in the model (see
text). All
quantities are in MeV.

\noindent {\bf Table IV} Same as Table II, but using a
Bethe-Weizsacker like mass formula. Values obtained for
incompressibility $K_{\infty}$ using these in Eq.(11) together with
the values from Table III for the corresponding set of data are given.
All quantities are in MeV.

\noindent {\bf Table V} Values of $K_{\infty}$ obtained with
inclusion/non-inclusion of higher-order effects like curvature
$a_{cv}^L$, proton-form-factor (p.f.f.)
and exchange Coulomb.
INM4 stands for INM model mass fit with 4 parameters, namely $a_v^I$,
$a_s^I$, $a_a^I$ and $a_C^I$. And, BW6 stands for BW model fit with 6
parameters, the two additional parameters in this case are the
pairing $\Delta$ and surface-symmetry $a_{ss}^L$ terms.

\vfill
\newpage
\centerline {\bf Table I}
\vspace {0.5in}
\begin{center}
\begin{tabular}{|c|c|c|c|}
\hline
\multicolumn{1}{|c|}{Parameters} &
\multicolumn{1}{|c|}{Exact Values} &
\multicolumn{1}{|c|}{INM} &
\multicolumn{1}{|c|}{BW} \\
\hline
$a_v$&15.87&15.925&14.769\\
$a_C$&0.757&0.7360&0.6945\\
$a_s$&17.3&18.10&11.15\\
$a_{a}$&27.0&29.80&25.41\\
$a_{ss}$&-16.0&-31.37&-17.77\\
$a_{cv}$&11.1&5.06&16.43\\
\hline
\end{tabular}
\end{center}
\vskip 1.0 true cm
\centerline {\bf Table II}
\vspace {0.5in}
\begin{center}
\begin{tabular}{|c|c|c|c|c|}
\hline
\multicolumn{1}{|c|}{No. of nuclei } &
\multicolumn{1}{|c|}{$a_v^I$} &
\multicolumn{1}{|c|}{$a_C^I$} &
\multicolumn{1}{|c|}{$a_s^I$} &
\multicolumn{1}{|c|}{$a_{a}^I$} \\
\hline
1085&16.101&0.7592&19.18&24.65\\
1191&16.115&0.7596&19.25&24.56\\
1252&16.112&0.7589&19.23&24.66\\
1294&16.096&0.7572&19.23&24.32\\
1371&16.108&0.7593&19.27&24.06\\
\hline
\end{tabular}
\end{center}
\newpage
\centerline {\bf Table III}
\vspace {0.5in}
\begin{center}
\begin{tabular}{|c|c|c|c|c|c|c|c|c|}
\hline
\multicolumn{1}{|c|}{No. of Para.} &
\multicolumn{1}{|c|}{$a_v^L$} &
\multicolumn{1}{|c|}{$a_s^L$} &
\multicolumn{1}{|c|}{$a_a^L$} &
\multicolumn{1}{|c|}{$a_C^L$} &
\multicolumn{1}{|c|}{$\Delta$} &
\multicolumn{1}{|c|}{$a_{ss}^L$} &
\multicolumn{1}{|c|}{$a_{cv}^L$} &
\multicolumn{1}{|c|}{$a_{gc}^L$} \\
\hline
5&-15.80&18.4&23.0&0.733&11.9& & & \\
6&-15.64&18.2&26.6&0.713&11.2&-22.3& & \\
7&-15.48&17.2&26.3&0.707&11.2&-21.9&1.5& \\
8&-12.66&-10.6&21.7&0.651&10.7&-7.9&102.7&-127.3\\
\hline
\end{tabular}
\end{center}
\vskip 1.0 true cm
\centerline {\bf Table IV}
\vspace {0.5in}
\begin{center}
\begin{tabular}{|c|c|c|c|}
\hline
\multicolumn{1}{|c|}{No. of nuclei} &
\multicolumn{1}{|c|}{$a_v^L$} &
\multicolumn{1}{|c|}{$a_C^L$} &
\multicolumn{1}{|c|}{$K_{\infty}$} \\
\hline
1085&15.648&0.7142&291\\
1191&15.640&0.7134&290\\
1252&15.651&0.7141&297\\
1294&15.634&0.7128&303\\
1371&15.635&0.7131&288\\
\hline
\end{tabular}
\end{center}
\newpage
\centerline {\bf Table V}
\vspace {0.5in}
\begin{center}
\begin{tabular}{|c|c|c|}
\hline
\multicolumn{1}{|c|}{} &
\multicolumn{1}{|c|}{Model Set} &
\multicolumn{1}{|c|}{$K_{\infty}$} \\
\hline
{\rm  With Exc. Coul.} & INM4 $\&$ BW6 & 288 \\
 & INM4 $\&$ BW6+$a_{cv}^L$ & 302 \\
 &  INM4+p.f.f. $\&$ BW6 + p.f.f. &  326 \\
 & & \\
 & & \\
{\rm  Without Exc. Coul.} & INM4 $\&$ BW6 & 303 \\
 & INM4 $\&$ BW6+$a_{cv}^L$ & 309 \\
 &  INM4+p.f.f. $\&$ BW6 + p.f.f. &  330 \\
\hline
\end{tabular}
\end{center}
\end{document}